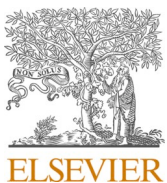
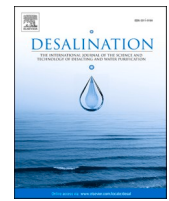

# From nature to engineering: Bio-inspired spacer solutions for membrane distillation

Alaa Adel Ibrahim [*], Stephan Leyer

*Department of Engineering, University of Luxembourg, 6 Rue Richard Coudenhove-Kalergi, 1359 Kirchberg, Luxembourg*

## HIGHLIGHTS

- New spacer designs are inspired by alligator osteoderms.
- Mixing enhancement for membrane distillation.
- New designs outperform conventional spacers in pressure drop, heat, and mass transfer.

ABSTRACT

Biomimetic design principles, inspired by the structure of alligator osteoderms, are employed to provide innovative solutions for enhancing membrane distillation system performance. A novel spacer design for membrane distillation (MD) systems is introduced, incorporating the unique structural features of these natural formations. Three-dimensional computational fluid dynamics (CFD) simulations are performed to investigate nine new spacer configurations, with particular emphasis placed on enhancing mixing efficiency within the feed channel, which serves as the primary focus of this research. Results demonstrate that the novel spacer configurations exhibit superior performance, achieving up to a fivefold increase in mixing efficiency relative to the baseline of an empty channel. Furthermore, the new spacers are assessed and contrasted with two types of conventional spacers and the empty channel configuration with respect to pressure drop, thermal performance, and mass transfer properties.

## 1. Introduction

Water is a critical resource found on Earth's surface, playing a fundamental role in sustaining life. However, ensuring reliable access to safe drinking water presents significant challenges. In the last century, rapid population growth, pollution, and climate change have become significant drivers that intensify pressure on water resources, ultimately leading to the emergence of water scarcity [1] Alternative water sources, such as desalinated seawater and brackish water, are necessary to address the existing gaps in water supply.

Within this framework, membrane distillation (MD) represents a promising technology that began in the 1960s with Bodell's patent for producing potable water from aqueous mixtures [2]. After a period of limited exploration, was revitalized in the 1980s, driven by significant advancements in membrane technology, particularly with polytetrafluoroethylene (PTFE) membranes [3]. Since that time, MD technology has experienced considerable improvements and has been thoroughly investigated for applications in desalination, wastewater treatment, resource recovery, and food processing industries [4].

Membrane distillation (MD) is a hybrid membrane-thermal separation process that relies on thermally driven vapor transport from the heated saline side through a non-wetted hydrophobic membrane, where the water vapor condenses on the cooler permeate side. This process effectively separates water from saline solutions while rejecting non-volatile solutes, with the driving force being vapor pressure gradient, generated by the temperature difference between the hot saline side and cold permeate side [4]. There are four primary configurations for MD, which are distinguished by the design of the permeate side. Direct contact membrane distillation is the most popular and simplest arrangement (DCMD). An air gap is created between the membrane and the condensate surface using air-gap membrane distillation (AGMD). While sweeping gas membrane distillation (SGMD) employs an inert gas on the permeate side to move vapor that condenses outside the membrane module, vacuum membrane distillation (VMD) applies vacuum pressure on the permeate side to improve the pressure gradient.

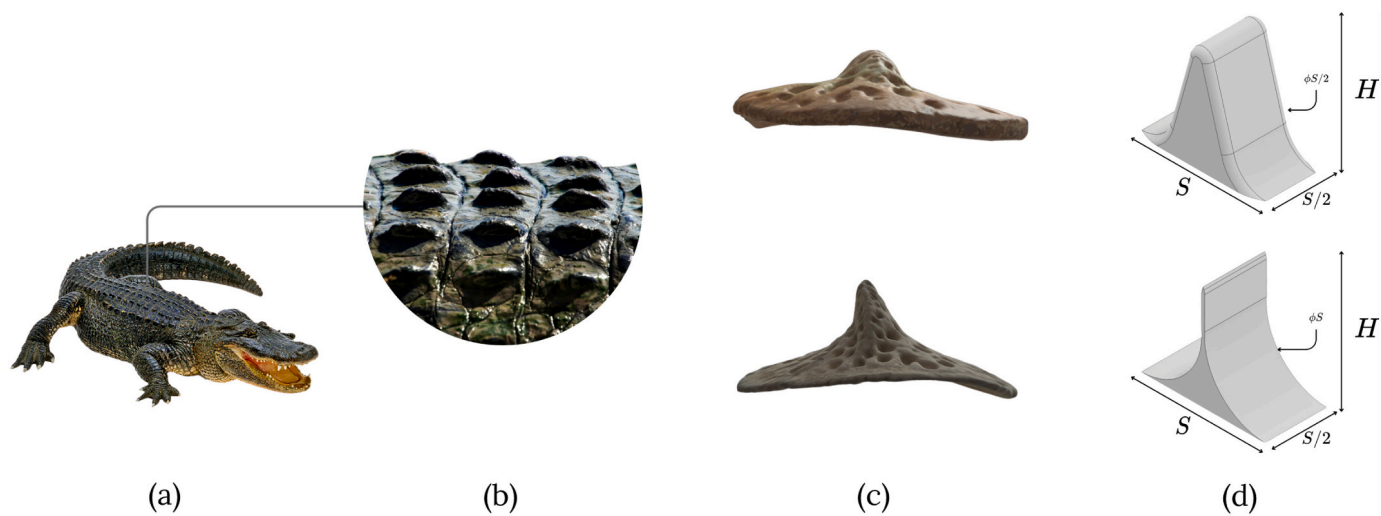

**Fig. 1.** Biomimetic module design for MD systems. (a) Alligator, (b) Close-up of alligator's back surface, (c) Samples of alligator osteoderms, (d) Proposed spacer designs, with Design 1 shown at the top and Design 2 at the bottom.

Although MD is still in the developmental phase compared to well-established membrane technologies like reverse osmosis (RO), it has achieved significant progress and attracted interest for its potential in the treatment of saline water and wastewater. MD stands out for its energy-efficient operation and cost-effectiveness compared to conventional distillation systems. Its ability to operate at low temperatures ($40^0$–80 °C) allows the use of low-grade industrial waste heat, solar thermal energy, and geothermal energy, further enhancing its energy efficiency. Moreover, MD exhibits a particular aptitude for the treatment of high-salinity brines and can be effectively scaled for decentralized applications in remote localities. These characteristics position MD as a promising alternative in the evolving landscape of water treatment technologies. Despite its potential advantages, MD faces several limitations. The main challenges include membrane fouling, wetting, and the occurrence of temperature and concentration polarization near the membrane interface. To tackle these limitations, researchers have extensively investigated the use of membrane spacers in MD systems. Originally used just to keep the membrane in place spacers are used to enhance mixing, reduce temperature polarization, and minimize fouling, ultimately improving permeate flux [5]. Their integration offers a practical approach to overcoming the inherent challenges of MD technology, aiming to enhance its overall efficiency and applicability in water treatment processes.

Extensive research, utilizing experimental, numerical, and theoretical methods, has been undertaken to clarify the function of spacers in MD systems [6]. The majority of these investigations have utilized numerical methods, with a particular emphasis on Computational Fluid Dynamics (CFD), to simulate MD systems incorporating spacers [7]. These computational techniques have been crucial in the optimization of spacer characteristics, including their geometry, arrangement, and material [8,9]. The primary objective of these optimizations has been to improve the processes of heat and mass transfer by mitigating the impacts of temperature polarization and concentration polarization.

While such technical approaches have advanced spacer design, nature offers a wealthy source of inspiration for solving engineering challenges through billions of years of evolutionary optimization. However, bio-inspired designs remain underutilized in membrane processes, with only limited applications thus far. Notable examples include honeycomb-based configurations that improved shear distribution and reduced fouling in forward osmosis systems [10], and bird V-formation-inspired modules that achieved remarkable 338 % flux enhancement in ultrapermeable membrane (UPM) systems for seawater desalination [11]. Building on this potential, we propose a novel spacer design inspired by alligator osteoderm structure, which exhibits naturally optimized hydrodynamic and structural properties.

To evaluate this biomimetic concept, work introduces a three-dimensional CFD investigation to evaluate the performance of the proposed spacer design. Inspired by the structure of alligator osteoderms, the design includes multiple configurations adaptable to different MD systems. Focusing on mixing performance within the feed/hot channel, our results demonstrate that these novel spacer configurations achieve up to 5 times enhancement in mixing efficiency compared to an empty channel baseline. This considerable improvement highlights the promising potential of nature-inspired solutions in advancing membrane technology.

## 2. Methodology

### 2.1. Bio-inspired spacers design

The alligator Fig. 1a, a well-preserved living fossil, exhibits remarkable adaptation through its dorsal armor Fig. 1b. This dermal shield comprises bony plates, known as osteoderms Fig. 1c, each featuring a central longitudinal keel. This structure serves a dual purpose, offering both protection and flexibility [12]. Beyond their protective function, the osteoderm keels promote the formation of a turbulent boundary layer and reduce flow separation by inducing localized turbulence along the alligator's body during swimming. These hydrodynamic effects contribute to drag reduction and efficient movement through water. Drawing inspiration from these biological defensive designs, characterized by irregular shapes and non-uniform keel heights, we developed two distinct spacer cell designs Fig. 1d. Both designs share identical length (S), depth (S/2), and height (H) corresponding to the channel height and is equal to 5 mm. The primary difference lies in the keel thickness, with the larger curvature diameter of ($\varphi$S) resulting in a thicker keel region. We designate the spacer design with the thicker keel as Design 1 (D1), and the thinner one as Design 2 (D2). This study examines the shape, arrangement, and orientation of these biomimetic spacers to evaluate their impact on the hydrodynamic characteristics within MD channel. Our investigation aims to demonstrate the potential benefits of biomimetic design in enhancing flow dynamics, not only in MD systems but also in the broader context of industrial mixing applications.

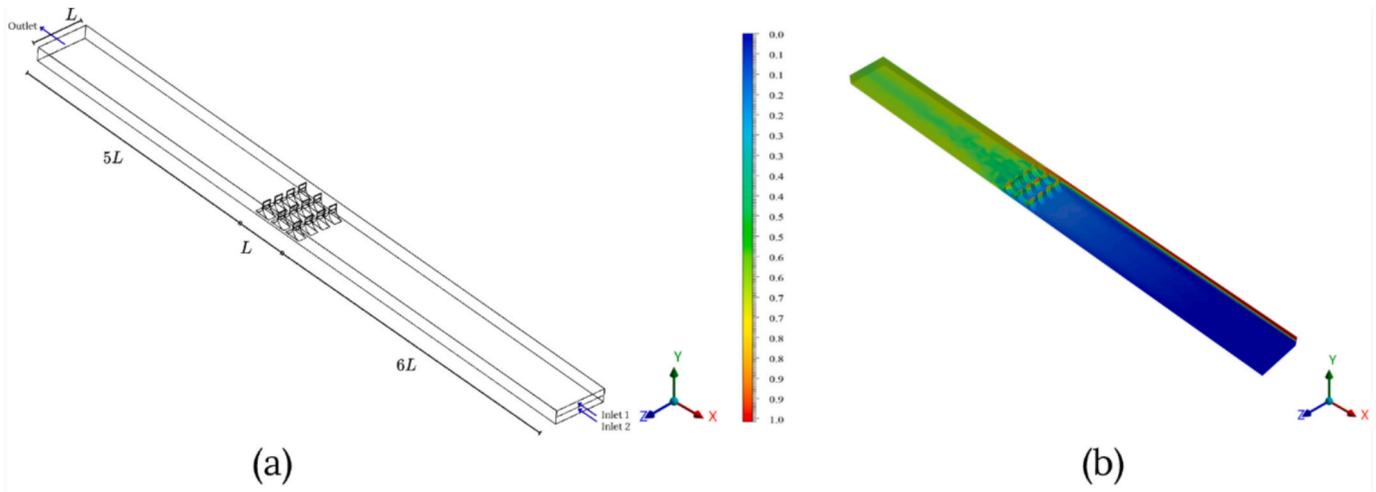

**Fig. 2.** Spacers oriented in the MD channel: (a) Computational domain and boundary conditions; L = 20 mm, (b) Dye mass fraction contour for the same spacer configuration; Design 1, FD2, inline configuration is used above.

**Table 1**
Novel Spacers Design Variables.

| Spacer name | Cell design | Configuration | Orientation ($\theta°$) | Void fraction ($\varepsilon$) |
| --- | --- | --- | --- | --- |
| D1-FD1-Inline | | Inline | 0° | 2.16E-02 |
| D1-FD1-Staggered | | Staggered | 0° | 1.56E-02 |
| D1-FD2-Inline | | Inline | 90° | 2.16E-02 |
| D1-FD2-Staggered | | Staggered | 90° | 1.56E-02 |
| D2-FD1-Inline | | Inline | 0° | 6.89E-03 |
| D2-FD1-Staggered | | Staggered | 0° | 5.74E-03 |
| D2-FD2-Inline | | Inline | 90° | 6.89E-03 |
| D2-FD2-Staggered | | Staggered | 90° | 5.74E-03 |
| DL-FD2 | | Up and down | 90° | 6.89E-03 |

### 2.2. Computational domain

Consistent with our previous research [13], we constructed a computational domain of length 6 L (See Fig. 2a). The spacer configurations were implemented above the membrane surface, positioned downstream of the fully developed flow region. This approach allows for a focused investigation of the mixing performance and pressure drop induced by each studied configuration. The working fluids employed in this study were water along with a dye tracer. We examined a range of Reynolds numbers from 80 to 1990, corresponding to inlet flow velocities between 0.01 m/s to 0.25 m/s.

Nine distinct configurations, shown in Table 1, are evaluated in this study based on the two proposed spacer designs. Both inline and staggered arrangements were tested using two flow directions (FD): FD1, where the angle between the flow direction and the z-axis ($\theta$) is 0°, and FD2, where $\theta$ is 90°. Additionally, a dual level (DL) configuration was tested, featuring alternating upward and downward cells based on Design 2 (D2). All nine configurations were systematically tested across the velocity range under investigation. To provide a comprehensive analysis, we compared the results from these novel configurations with those obtained from an empty channel case and a conventional spacer design, characterized by a cylindrical net-shaped structure. This comparative approach allows for a comprehensive assessment of the proposed biomimetic spacers' performance relative to standard MD configurations. In Fig. 2b, the dye mass fraction contour demonstrates the notable influence of the (D1-FD2-Inline) spacer, with further discussion provided in the results section.

### 2.3. Numerical simulations

The research employed 3D computational fluid dynamics (CFD) analysis using ANSYS Fluent 2022 R2 to solve Navier-Stokes equations. For analyzing flow patterns around spacer designs, a transient turbulent model was used, while the empty channel setup used a laminar flow model. The same numerical approach as outlined in our previous work [13] is adopted here. The computational process, shown in Fig. 3, followed several steps for spacer optimization. First, spacers were designed with Inventor software, followed by geometry creation featuring two models: split inlet boundaries (water and dye inlets for mixing studies) and non-split inlet (for heat transfer analysis). Mesh generation utilized a 0.0004 m mesh size, determined through sensitivity analysis in our previous work [13]. The simulation was configured with specific fluid properties (Table 2) and boundary conditions, including inlet velocities of 0.01–0.25 m/s for both dye and water, with zero outlet pressure. The study employed species transport and the SST k-ω turbulence model. Results underwent convergence testing and analysis of key parameters including mixing index, dye mass fraction, pressure drop, and Sherwood number. This process resulted in nine different spacer configurations with two different cell designs, were systematically evaluated. Their performance was compared to that of an empty channel and two conventional spacer designs composed of circular cross-section fibers: one with fibers oriented orthogonally at 90°, and the other arranged at a 45° inclination relative to the flow direction [13]. These conventional spacers are hereafter referred to as CS-90 and CS-45, respectively. Model validation was achieved through comparison with both numerical and experimental data reported in [14] and [15], demonstrating good agreement, as discussed in our previous study [13].

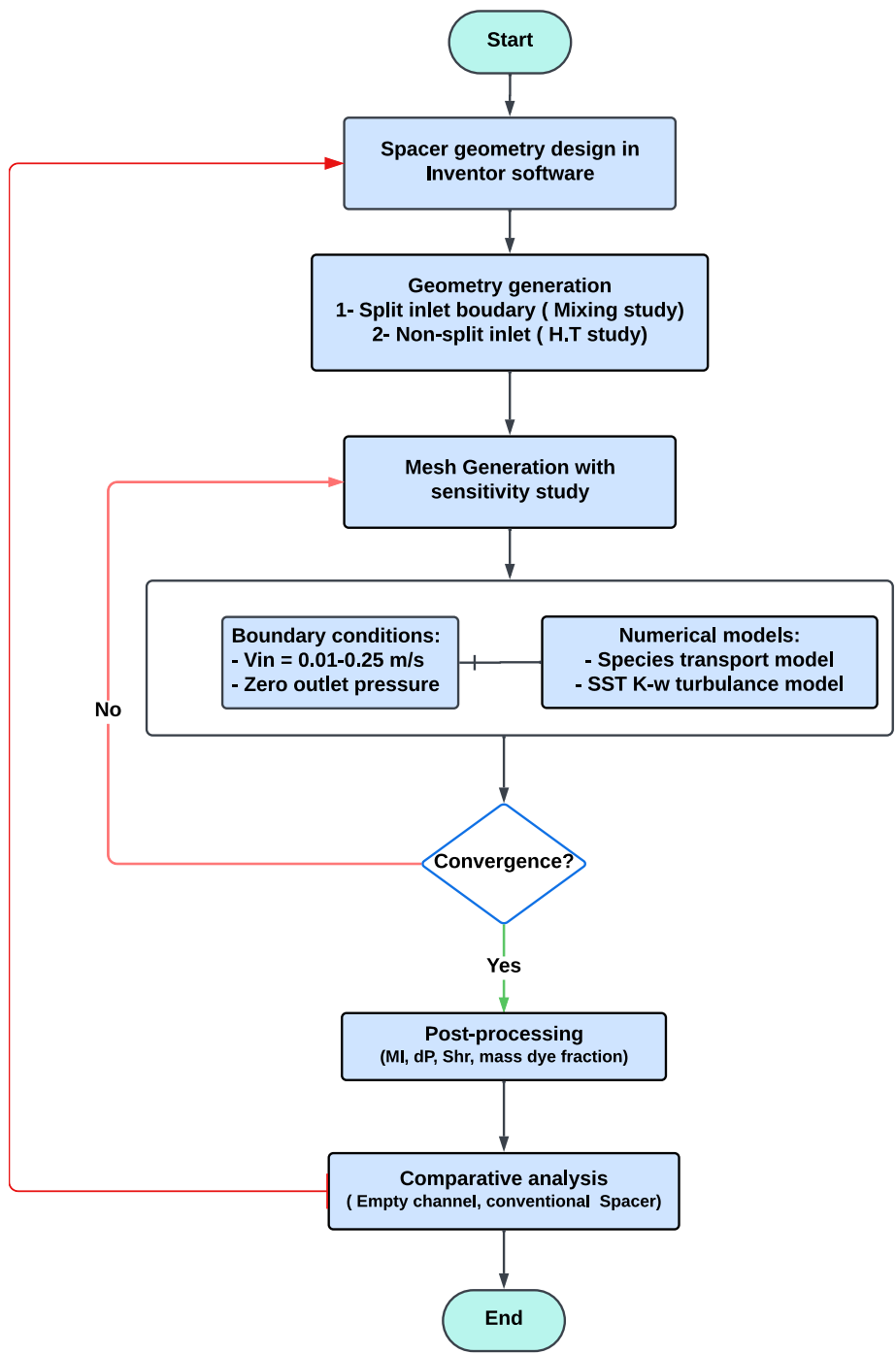

**Fig. 3.** Overview of the numerical framework.

**Table 2**
Physical characteristics of the channel feed, membrane, and bio-spacers.

| Material Property | Units | Water & Dye | PVDF | Copper |
|---|---|---|---|---|
| Density, $\rho$ | kg/m$^3$ | 998.2 | 1176 | 8978 |
| Specific heat, $c_p$ | kJ/kg.K | 4.182 | 1.325 | 0.381 |
| Conductivity, k | W/m.K | 0.6 | 0.2622 | 387.6 |
| Dynamic viscosity, $\mu$ | kg/m·s | 0.001003 | – | – |
| Mass diffusivity, D | m$^2$/s | 2.88e − 50 | – | – |

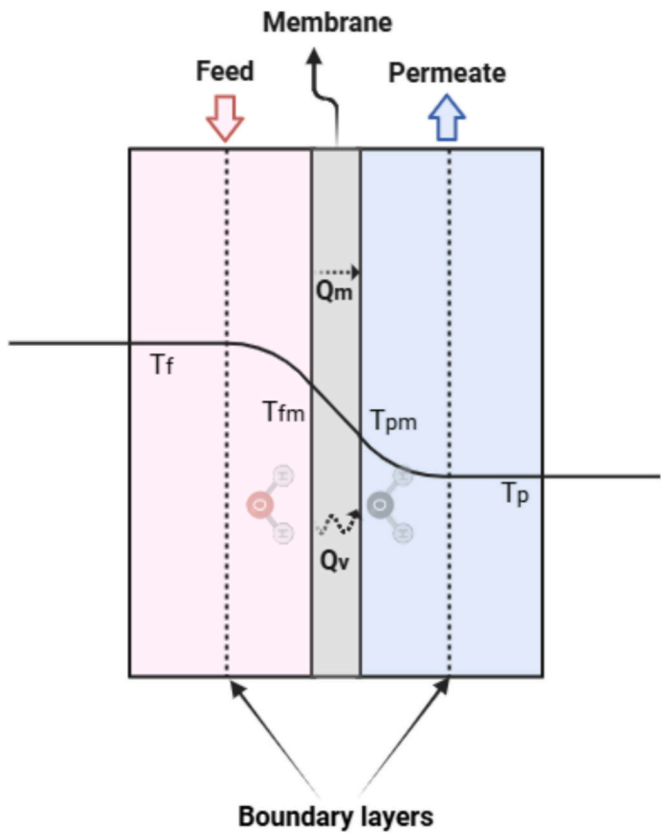

**Fig. 4.** Illustration of mass and heat transfer mechanisms within membrane distillation systems.

## 2.4. Spacer performance metrics

Spacer performance in MD systems is evaluated through multiple indicators that quantify their ability to enhance flow mixing, mass and heat transfer, and minimize polarization effects in feed channels.

### 2.4.1. Flow mixing assessment

In this context, mixing phenomena was assessed through both qualitative and quantitative approaches. The numerical domain was configured such that both dye and water entered at equal velocities through equivalent cross-sectional areas. By introducing turbulence through the proposed bio-inspired spacers, the homogeneity of the two fluids qualitatively was assessed by visualizing dye mass fraction profile on a plane positioned at the terminus of the spacer, perpendicular to the flow direction (see Fig. 5). Quantitatively, the degree of mixing could be assessed using the mixing index (MI) or the coefficient of variation (CV) [16–19], which are statistical measure calculated based on analysis of samples taken at the channel's outflow cross-section.

The mixing index (MI) is calculated using the coefficient of variation (CV), which acts as an inverse indicator of mixing quality. The MI is determined by the following formula [16]:

$$MI = 1 - \frac{\sigma}{\overline{C}} = 1 - CV \qquad (1)$$

where

$$\sigma = \sqrt{\frac{1}{N}\sum_{i=1}^{N}(C_i - \overline{C})^2} \qquad (2)$$

In this formulation, $\sigma$ is the Standard deviation of the dye mass fraction, $\overline{C}$ is the Mean dye mass fraction, N represents the total number of sampling points at the outflow cross-section where dye mass fraction is measured, and $C_i$ denotes the dye mass fraction at the i-th sampling point within the cross-section, where i ranges from 1 to $N$. The coefficient of variation (CV) is defined as the standard deviation divided by the mean ($CV = \frac{\sigma}{\overline{C}}$). This measure ranges from 0 to 1. The lower the mass fraction deviation from the mean value, the better the mixture homogeneity, thus indicating enhanced mixing. Consequently, as CV values get closer to 0, this indicates a high degree of fluid mixing. Conversely, when CV values near 1, this suggests minimal mixing has occurred within the fluid. [15].

The mixing index thus provides a quantitative metric for mixture with a positive correlation to mixing quality. When the mixing index equals 0, the fluids remain completely separate, whereas a value of 1 ($\sigma = 0$) indicates that the two species are fully mixed together.

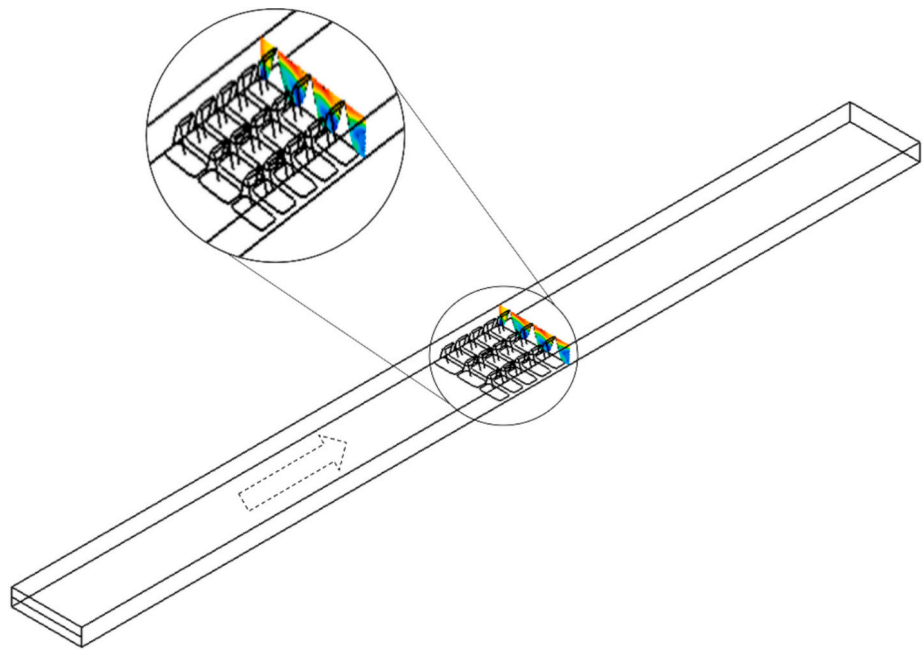

**Fig. 5.** The plane at the terminal of the spacers area and perpendicular to the flow direction.

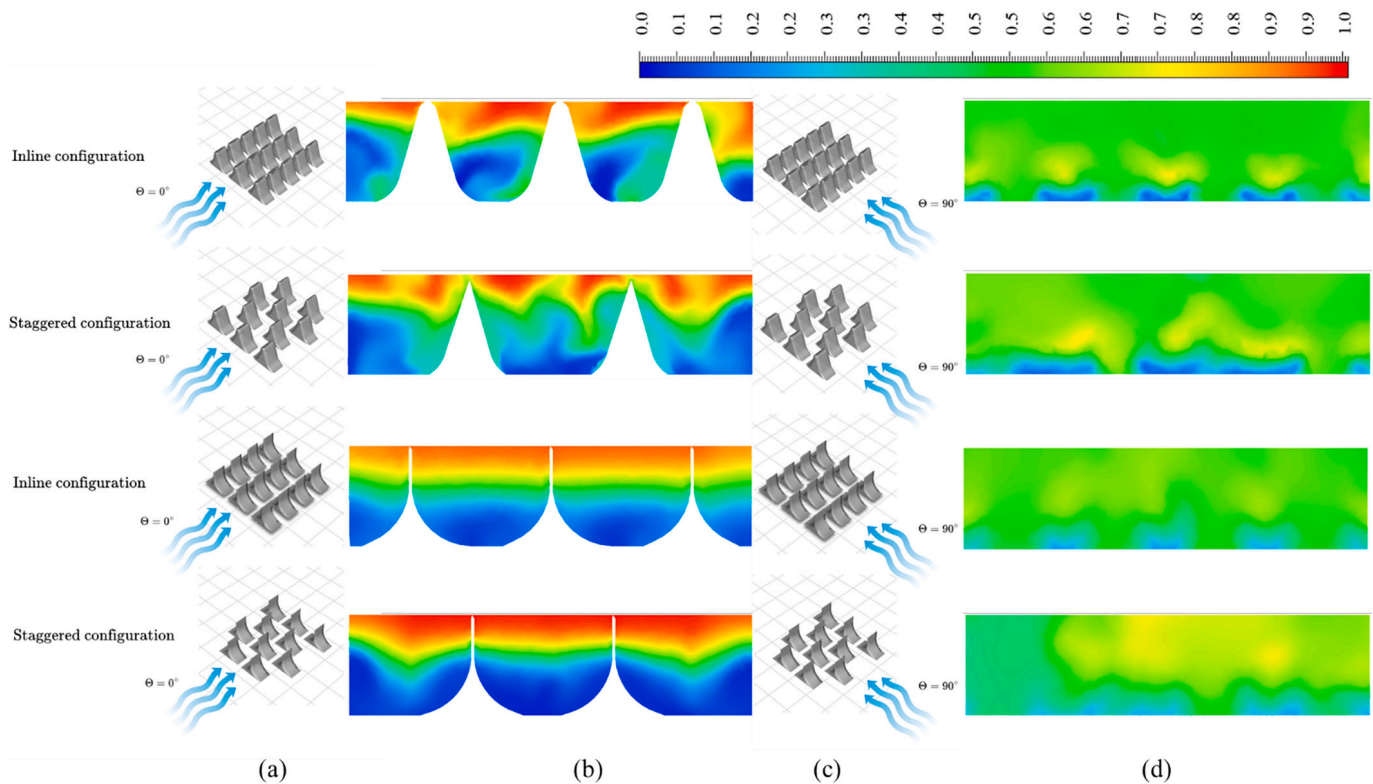

**Fig. 6.** Dye mass concentration distributions at fluid inlet velocity of 0.25 m/s on a cross-section at the spacer region's end. (a) Spacers configurations for FD1, (b) contour results for FD1, (c) Spacers configurations for FD2, (d) contour results for FD2.

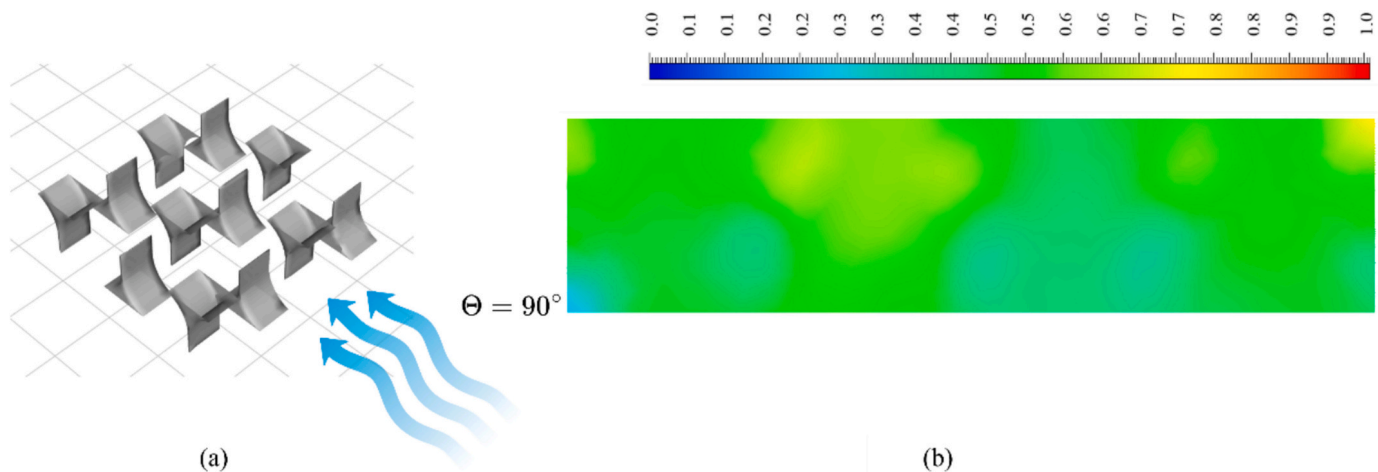

**Fig. 7.** Dye mass concentration distributions at fluid inlet velocity of 0.25 m/s on a cross-section at the spacer region's end. (a) Dual level (DL) configuration of upward and downward cells, (b) contour results for FD2.

*2.4.2. Mass and heat transfer characteristics*

In MD systems equipped with spacers, mass transfer occurs through the movement of substances across feed channels and across a hydrophobic, microporous membrane. This transfer process in MD is commonly described by Knudsen diffusion, Poiseuille flow, molecular diffusion, or combinations of these mechanisms [20]. While most research on spacer-induced mass transfer employs simplified models with impermeable, non-absorptive wall assumptions, relatively few studies examine actual permeable membrane configurations [21]. The assessment of mass transfer effectiveness encompasses both experimental and theoretical methodologies [22,23].

In experimental studies, researchers quantify performance through permeate flux measurements [24–26], mass transfer coefficients, and the dimensionless Sherwood number (Sh). Theoretical investigations require the simultaneous solution of two fundamental equation sets: the Navier-Stokes equations governing fluid dynamics and the convection-diffusion equation describing mass transport. Both methods typically analyze results using the Sherwood number or mass transfer coefficient as performance indicators [27–32].

In this study, the Sherwood number (Sh) was used to analyze mass transfer characteristics. $Sh$ quantifies the ratio between convective and diffusive mass transfer processes, as defined in equation (Eq. (3)). The $Sh$ value varies based on membrane distillation (MD) configuration and feed flow parameters. We calculated $Sh$ using the Graetz-Lévêque

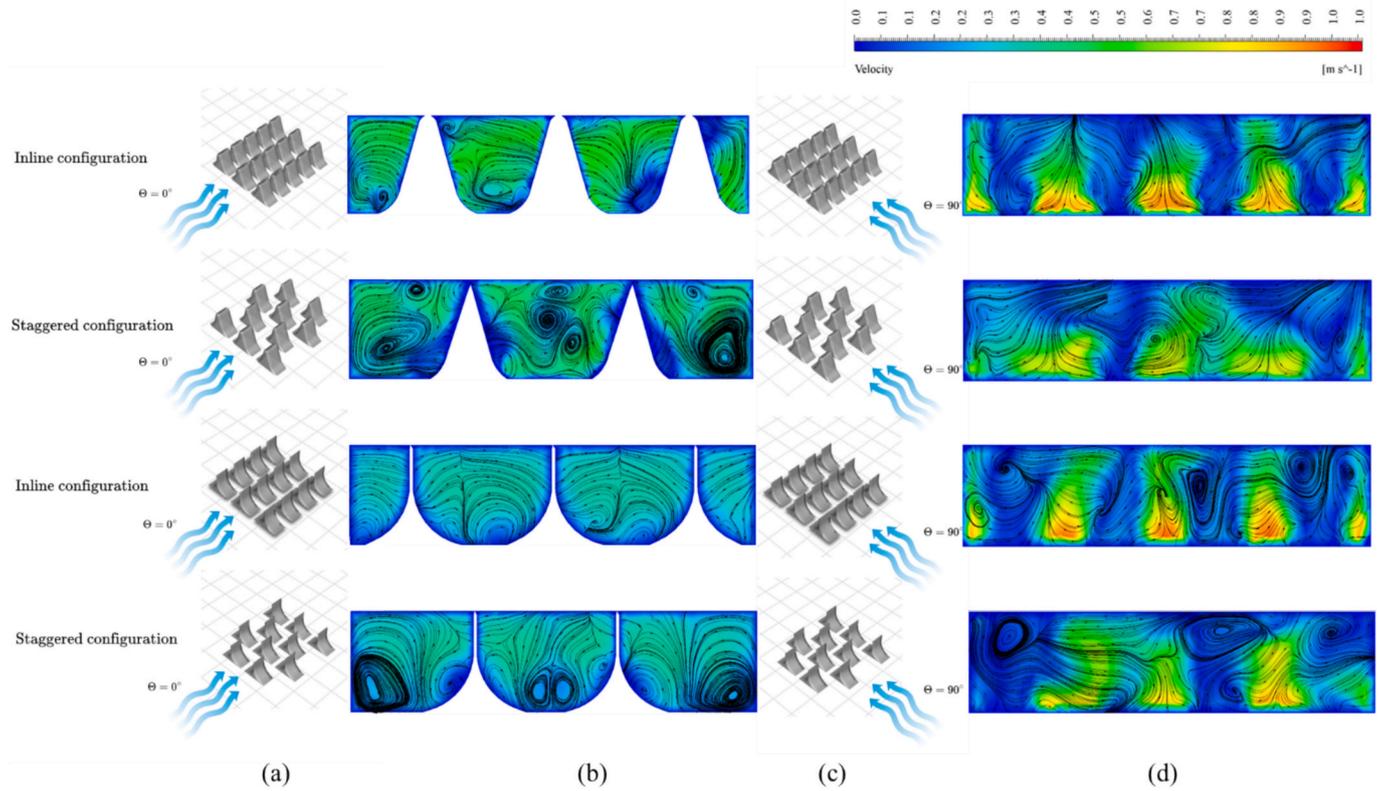

Fig. 8. Velocity contours with velocity streamlines at fluid inlet velocity of 0.25 m/s on a cross-section at the spacer region's end. (a) Spacers configurations for FD1, (b) contour results for FD1, (c) Spacers configurations for FD2, (d) contour results for FD2.

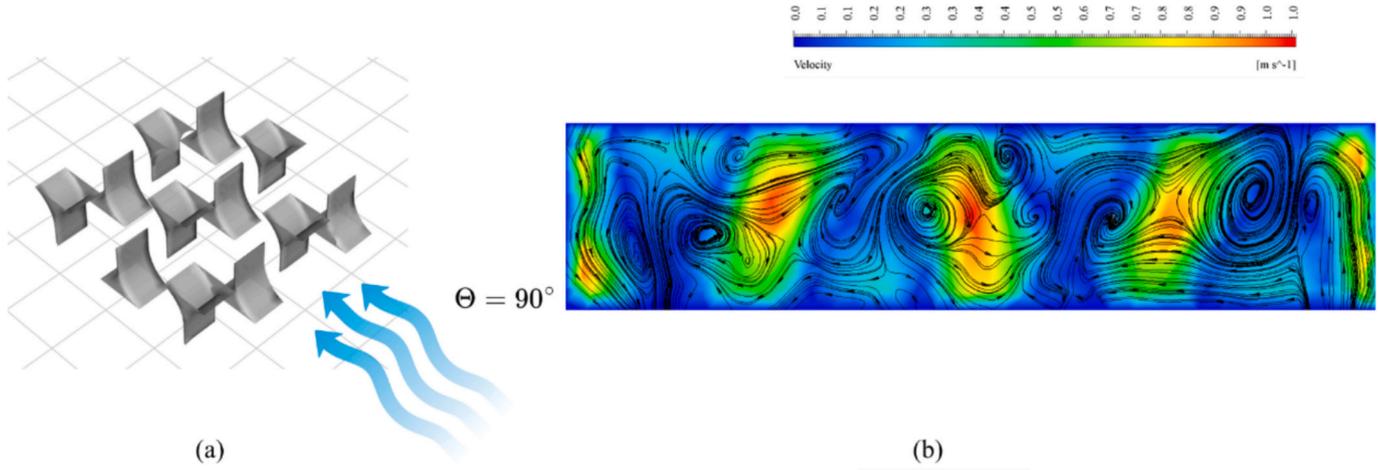

Fig. 9. Velocity contours with velocity streamlines at fluid inlet velocity of 0.25 m/s on a cross-section at the spacer region's end. (a) Dual level (DL) configuration of upward and downward cells, (b) contour results for FD2.

equation (Eq. (4)) [33].

$$Sh = \frac{k_m d_h}{D} \quad (3)$$

$$Sh = 1.86 \left( \frac{Re \times \mu \times d_h}{\rho \times D \times L} \right)^{0.33} \quad (4)$$

where $k_m$ and the mass transfer coefficient (m/s), $d_h$ is the hydraulic diameter of the channel (m), $Re$ is the Reynolds number, $\mu$ and $\rho$ are the feed flow dynamic viscosity (kg/m·s) and density (kg/m³) respectively, L is the feed channel length (m), and D represents diffusion coefficient using the experimentally measured value of $0.73 \times 10^9$ m²/s [34].

MD technology operates by utilizing the temperature gradient between the feed and permeate, which facilitates heat transfer from the hot side of the membrane to its cold side. This temperature difference triggers a phase change from liquid to vapor at the feed surface of the membrane. The heat transfer process takes place across three distinct regions: feed boundary layer, membrane ($Q_m$), and permeate boundary layer (see Fig. 4) [35]. As shown in Fig. 4, the bulk temperatures of hot feed and cold distillate are indicated by ($T_f$) and ($T_p$) respectively. The phenomenon of temperature polarization (TP) arises from the difference between membrane surface and bulk fluid temperatures, resulting in the feed-side membrane temperature decreasing to ($T_{fm}$) while the distillate-side membrane temperature rises to ($T_{pm}$). This phenomenon is quantified using the temperature polarization coefficient (TPC), calculated as the ratio of membrane temperature difference to bulk solutions temperature difference (Eq. (5)). A reduced temperature gradient at the membrane surface compared to the bulk reduces the driving force for water vapor flux [25].

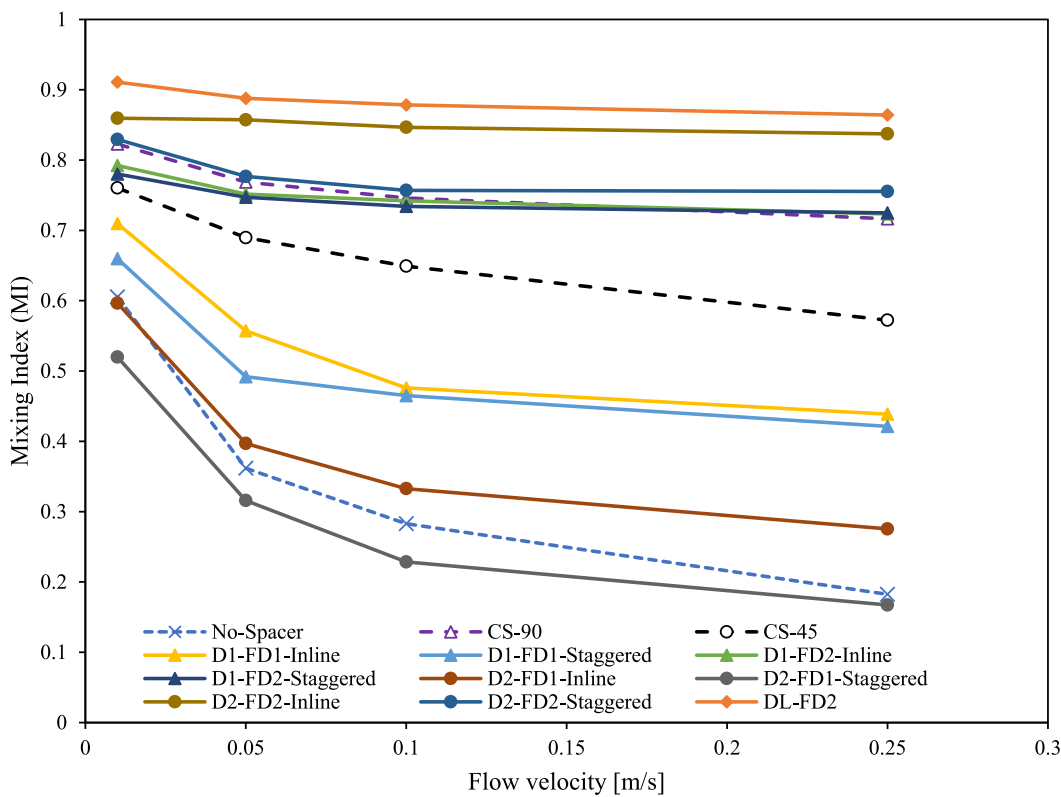

**Fig. 10.** Mixing Index (MI) for the proposed spacer configurations compared with empty and two conventional spacer types (CS-90 and CS-45), at inlet velocity range of: (0.01–0.25) m/s.

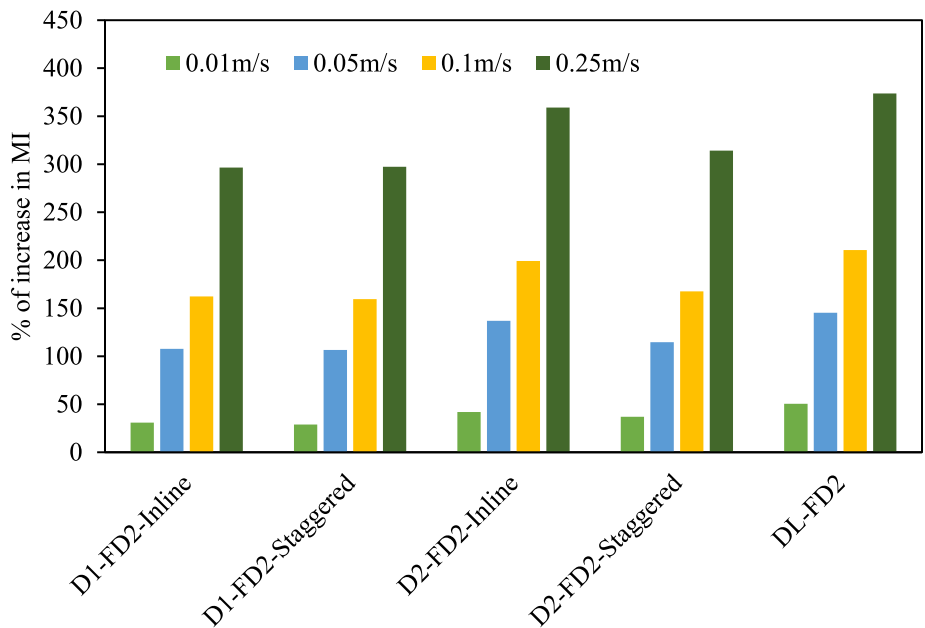

**Fig. 11.** Increase in Mixing Index (MI) for FD2 Configurations Compared to No-Spacer configuration.

$$TPC = \frac{T_{fm} - T_{pm}}{T_f - T_p} \quad (5)$$

In thermal processes like Membrane Distillation (MD), spacers enhance heat transfer by minimizing the thickness of the temperature boundary layer, thereby decreasing TP effect, and consequently increasing the driving force [25]. Experimentally, this is measured by monitoring temperatures at the membrane's feed ($T_{fm}$) and permeate

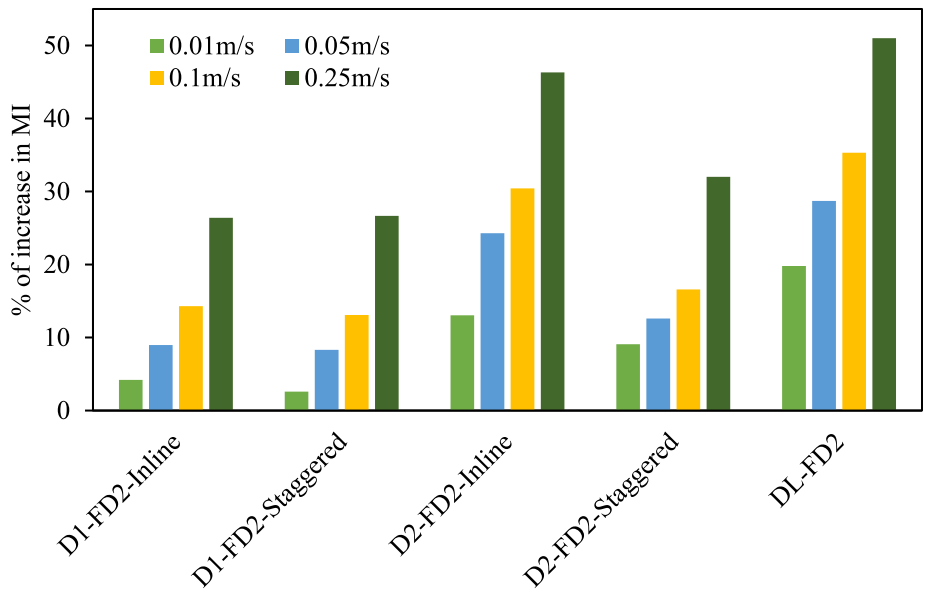

**Fig. 12.** Increase in Mixing Index (MI) for FD2 Configurations Compared to CS-45 configuration.

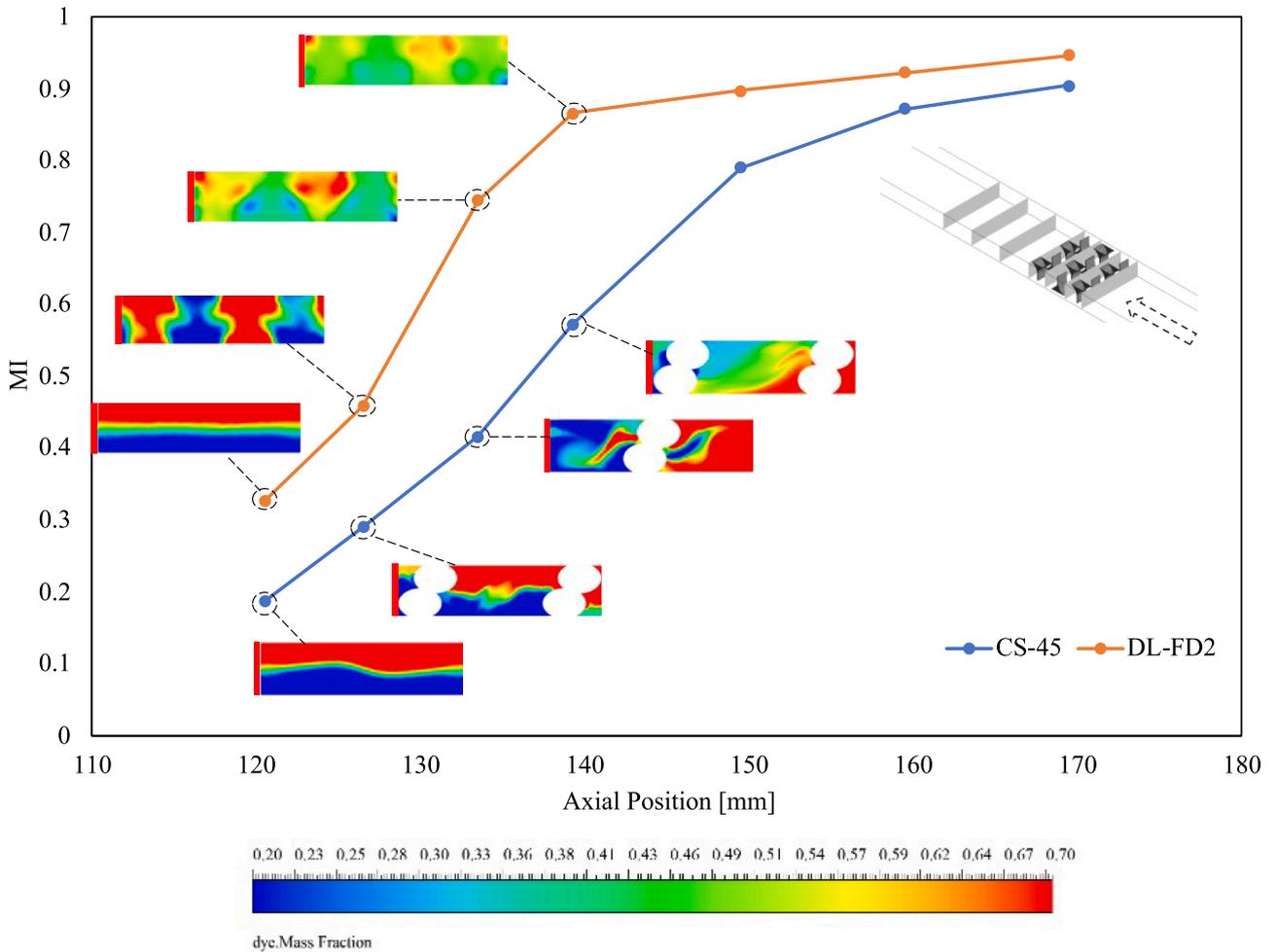

**Fig. 13.** Comparison of the Mixing index (MI) along the feed channel for DL-FD2 and CS-45 configuration, at fluid inlet velocity of 0.25 m/s.

sides ($T_{pm}$), along with water vapor pressure and flux. The theoretical approach entails solving continuity, momentum, and energy equations, and calculating the Nusselt number ($Nu$) to assess heat transfer improvements caused by fluid movement [21]. In this investigation, the average $Nu$ is numerically calculated to determine the heat transfer in the hot channel, where the feed temperature entering the channel is

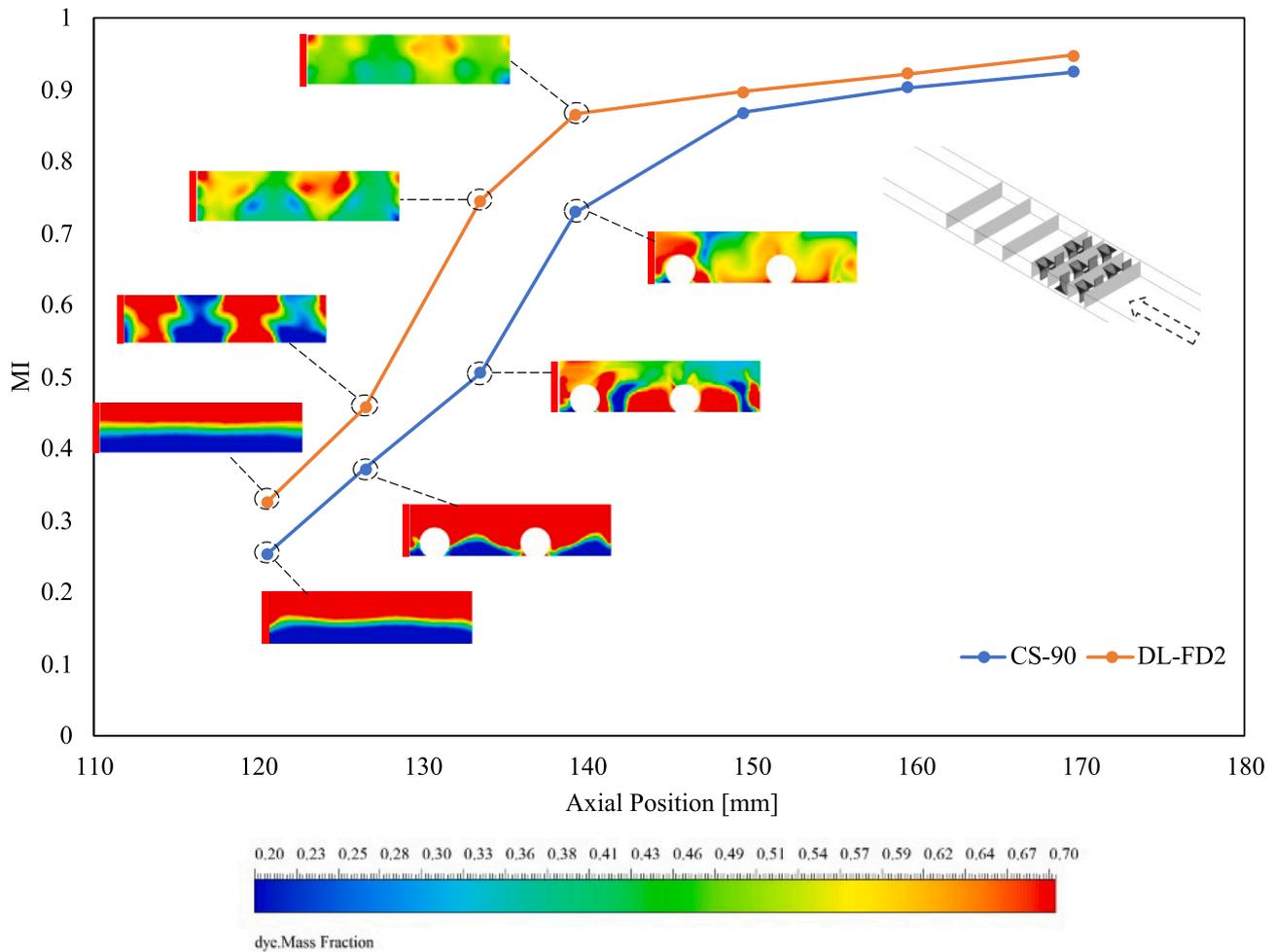

**Fig. 14.** Comparison of the Mixing index (MI) along the feed channel for DL-FD2 and CS-90 configuration, at fluid inlet velocity of 0.25 m/s.

57 °C and the membrane surface temperature is 27 °C. The calculation is performed using the following equation (Eq. (6)), where $h_{ave}$ represents the average heat transfer coefficient of the PVDF membrane surface, L denotes the characteristic length, and k represents the thermal conductivity of the fluid. Additionally, the spacer material is set to copper.

$$Nu = \frac{h_{ave} L}{k} \quad (6)$$

*2.4.3. Polarization and fouling behavior*

In contrast to other membrane technologies such as nanofiltration (NF) and reverse osmosis (RO), concentration polarization (CP) has a minimal effect in membrane distillation (MD) [36]. CP represents the accumulation of solutes close to the membrane surface creating concentration gradients, typically reduces separation efficiency in membrane processes. CP phenomenon is quantified using concentration polarization coefficient (CPC), calculated as concentration differences between the bulk ($C_f$) and the membrane surface ($C_{fm}$) values (Eq. (7)). Although MD is comparatively resistant to CP, membrane fouling remains a major obstacle to sustaining its long-term operational efficiency [37]. High CPC values and low TPC values are undesirable as they reduce the distillation driving force and encourage the formation of fouling [37].

$$CPC = \frac{C_{fm}}{C_f} \quad (7)$$

Studies indicate that effective feed spacer designs can help minimize fouling by increasing surface wall shear, which serves as an indicator for fouling reduction [38]. Numerous CFD modeling studies have established that enhanced turbulence disrupts the boundary layer close to the membrane to be disrupted (polarization effect) and raises shear stress [37–39]. Consequently, this study investigates shear stress distribution across the membrane surface to assess the impact of proposed spacer configurations.

## 3. Results and discussion

### 3.1. Mixing performance analysis

*3.1.1. Qualitative assessment of mixing performance*

The mixing performance of various biomimetic spacer configurations was assessed using dye mass fraction contour analysis. A total of nine configurations were examined, consisting of eight single-level designs and one dual-level arrangement.

Fig. 6 illustrates the dye mass fraction distributions for the single-level configurations, which incorporated two biomimetic spacer designs (D1 and D2) arranged in both staggered and inline orientations. Each configuration was further analyzed in two flow directions (FD1 and FD2). More uniform dye mass fraction contours indicate enhanced mixing efficiency. The results show that flow direction plays a critical role in mixing behavior. FD2-oriented configurations consistently exhibited greater homogeneity in dye mass fraction contours at the spacer region terminus compared to FD1-oriented configurations. This enhancement is attributed to the higher velocity regions in FD2-oriented cases, as demonstrated in Fig. 8.

The dual-level configuration, combining upward and downward cells oriented in FD2, achieved the highest degree of contour homogeneity among all tested configurations, as shown in Fig. 7, indicating optimal mixing performance. This configuration generated the highest velocity regions (Fig. 9) and the most intense vortex formation, which enhanced turbulence and mixing efficiency.

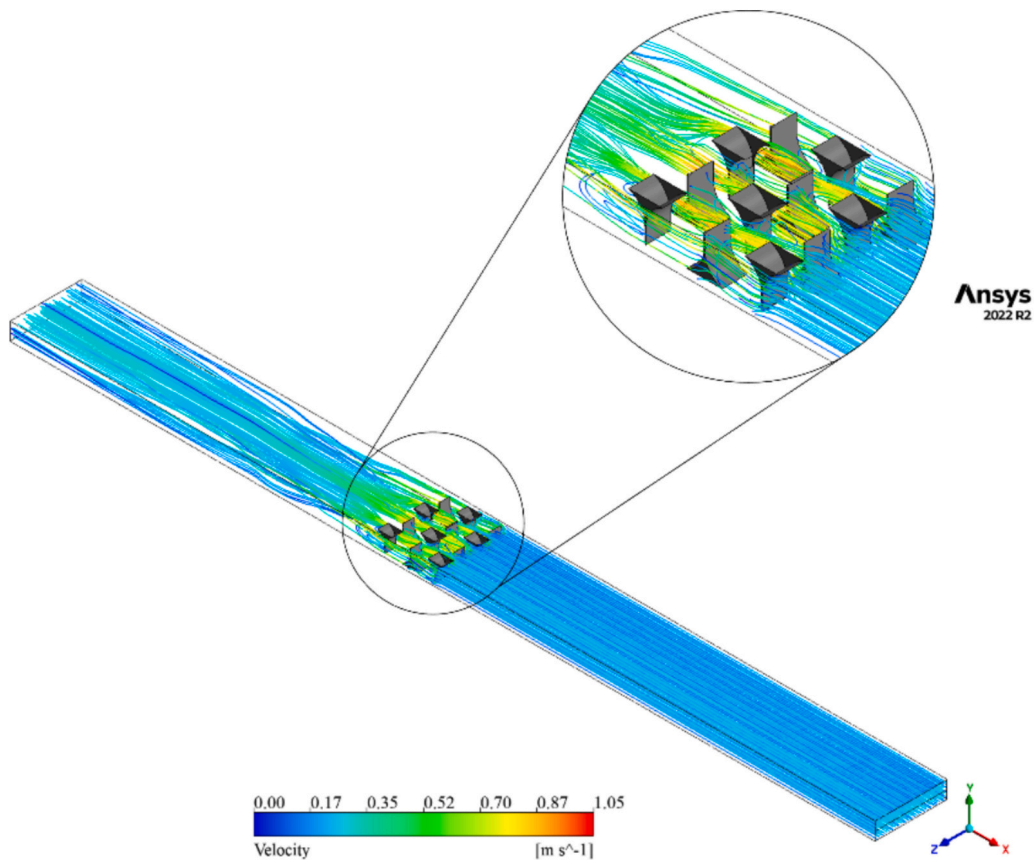

**Fig. 15.** 3D velocity streamlines in the feed channel with DL-FD2 spacer design implemented at an inlet flow velocity of 0.25 m/s.

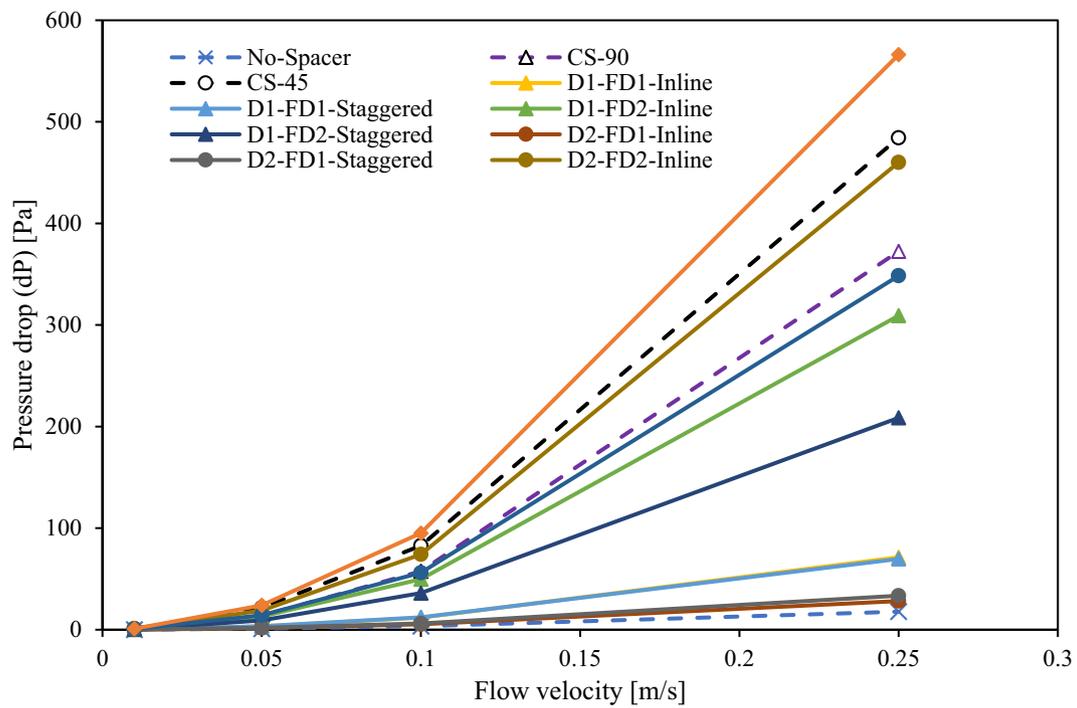

**Fig. 16.** Pressure drops against all the spacers configurations studied at inlet velocity range of: (0.01–0.25) m/s.

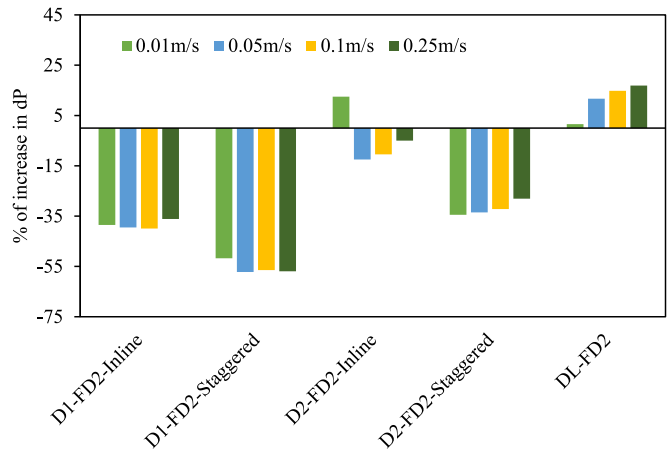

**Fig. 17.** Increase in Pressure drop (dP) for FD2 Configurations Compared to CS-45 Configuration.

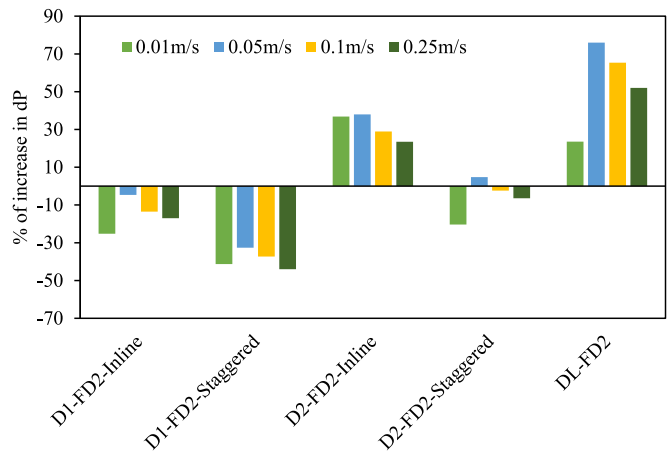

**Fig. 18.** Increase in Pressure drop (dP) for FD2 Configurations Compared to CS-90 Configuration.

*3.1.2. Quantitative assessment of mixing performance*

Fig. 10 demonstrates the mixing index (MI) values calculated for the new bio-inspired spacer configurations, allowing a comparison with both an empty (no-spacer) channel and two conventional spacer types, CS-90 and CS-45. Notably, five of these novel configurations showed improved mixing efficiency. The common factor of these five cases is the flow direction. Analysis of these high-performing configurations indicated that flow direction was the critical parameter influencing their performance. Specifically, Flow Direction 2 (FD2) consistently demonstrated superior mixing characteristics across all tested configurations, suggesting that spacer orientation relative to flow direction significantly impacts spacer functionality.

Fig. 11 and Fig. 12 provide a quantitative assessment of the performance enhancements achieved with FD2-aligned configurations compared to no-spacer and CS-45 spacers configurations. For the no-spacer setup, the data reveals a significant enhancement in mixing efficiency, with improvements ranging from nearly 300 % to 370 %, particularly in the dual-level (DL) configuration. When using CS-45 spacers, the results indicate a positive correlation between the inlet flow velocity and the percentage increase in the mixing index (MI). Design 1 (D1) exhibits an MI enhancement of approximately 26 % in both inline and staggered arrangements at the highest flow velocity. Design 2 (D2) demonstrates further improvements, with increases of 46 % and 32 %. The dual-level (DL) configuration yields the most substantial enhancement, achieving a 51 % increase in MI, with values approaching 0.9.

Given the promising mixing performance demonstrated by the DL configuration, a detailed analysis of the mixing index (MI) and dye mass fraction distributions along the feed channel. Seven perpendicular planes to the flow direction were analyzed: the first plane is positioned just before the spacers, the second plane immediately downstream of the first spacer filament, the third just downstream of the second filament, and the fourth at the terminal edge of the spacers. The final three planes are positioned downstream of the spacers, with a 10 mm separation between each. These measurements were compared with corresponding planes in CS-45 and CS-90 configurations, and the results are presented in Fig. 13 and Fig. 14. It can be deduced that effective mixing is achieved more quickly with the incorporation of the DL spacer configuration compared to the cylindrical spacers.

This integrated qualitative and quantitative analysis provided comprehensive insights into mixing dynamics through computational fluid dynamics (CFD) simulations, yielding both visual and numerical evidence of mixing performance under the specified flow conditions. The results demonstrate that spacer geometry, configuration, and flow direction alignment are critical parameters in optimizing mixing efficiency. The dual-level configuration emerged as the superior design, exhibiting both optimal dye concentration distribution patterns and the highest mixing index (MI ≈ 0.9), indicating exceptional mixing performance suitable for practical mixing-process applications [16]. Additionally, as shown in Fig. 15, the 3D velocity streamlines illustrate the flow behavior generated by the DL-FD2 spacer design at an inlet velocity of 0.25 m/s.

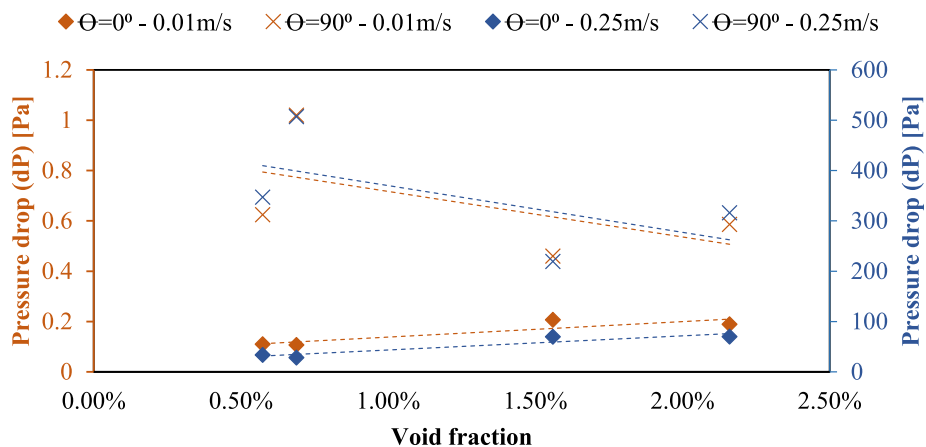

**Fig. 19.** Pressure drop vs. Spacer's void fraction; the dotted lines on the graph indicate the trendlines.

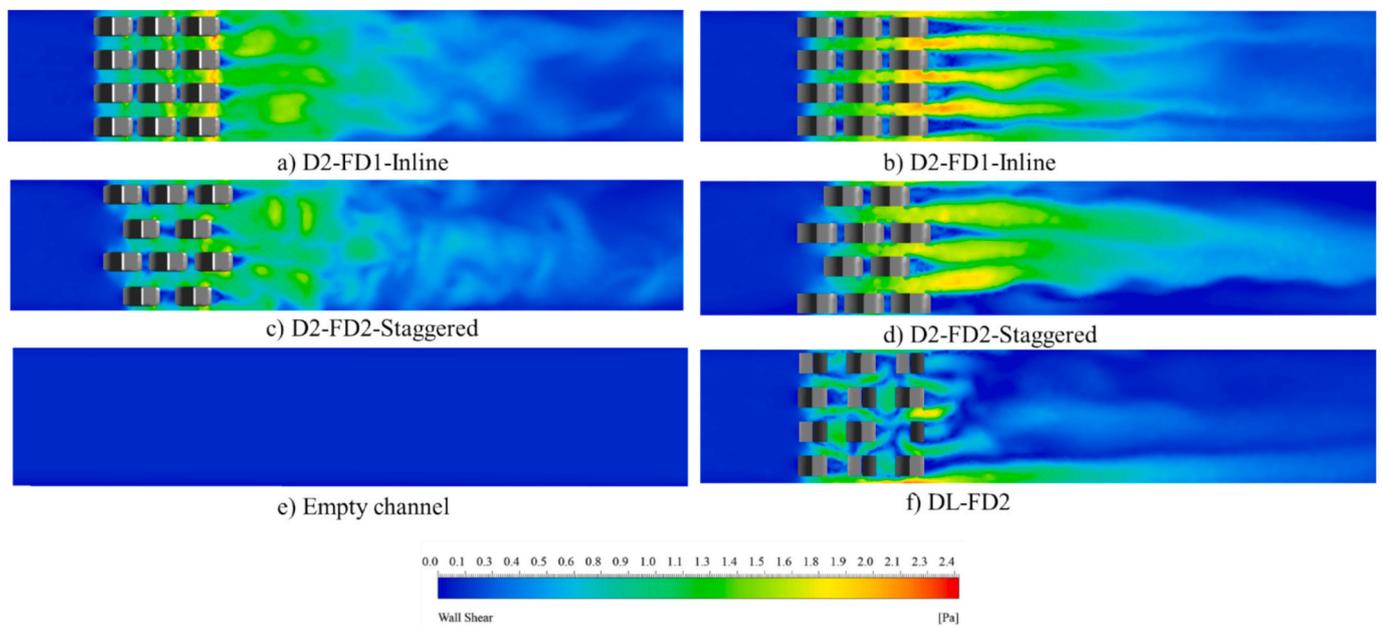

**Fig. 20.** Contours of wall shear stress over the membrane surface FD2 Configurations Compared to no-spacer configuration, inlet velocity = 0.25 m/s.

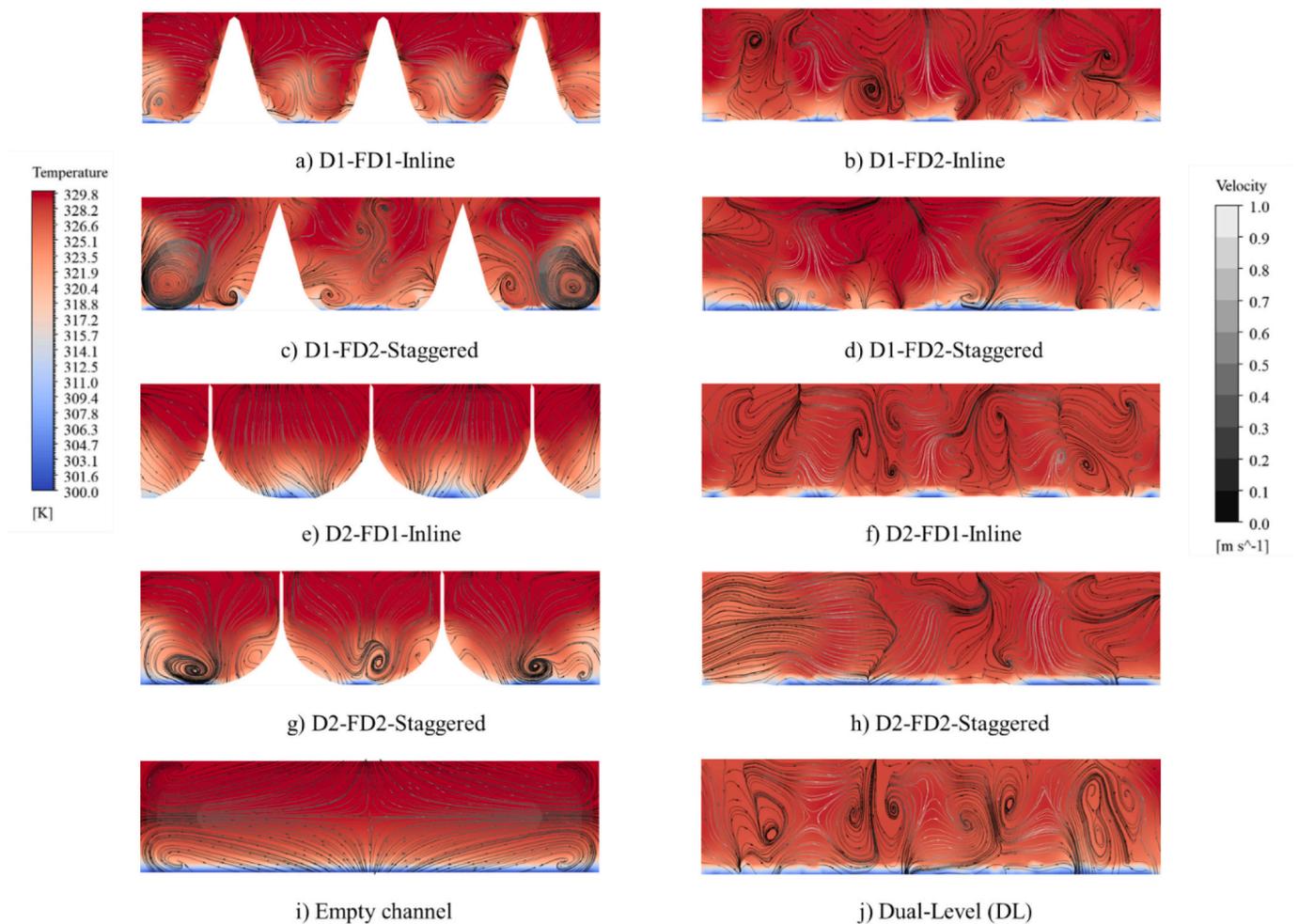

**Fig. 21.** Temperature contours with velocity streamlines at inlet flow velocity of 0.25 m/s.

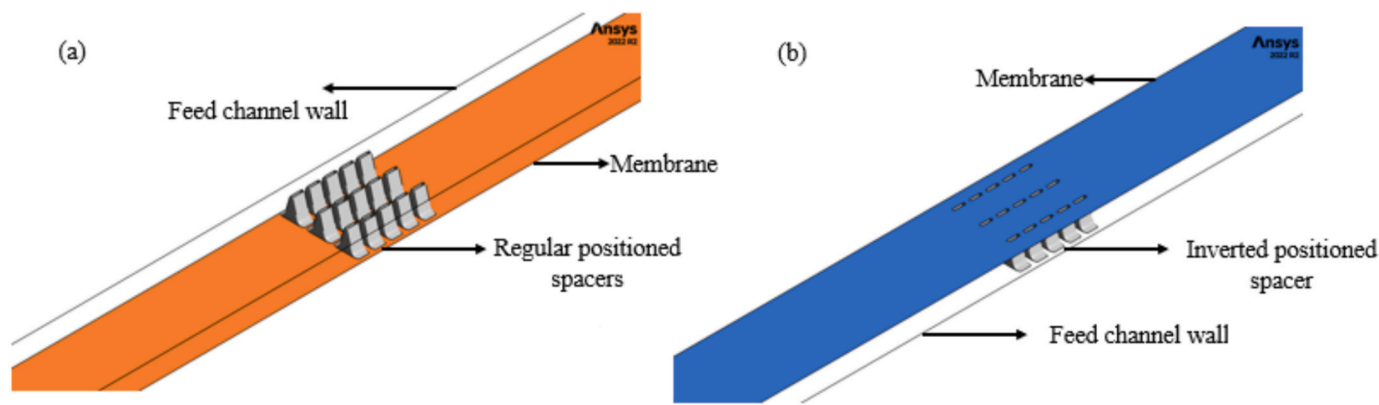

**Fig. 22.** Spacer Orientation Scenarios: (a) Regular orientation: spacers on the membrane surface, (b) Inverted orientation: spacers on the feed channel wall.

*3.2. Effect of the proposed bio-inspired spacers on pressure drop*

Fig. 16 illustrates the impact of adding and altering spacer configurations on the pressure drop, which directly influences the pumping power required. While the turbulence generated by the spacers enhances mixing, it also leads to an increase in pressure drop, presenting a critical challenge that requires optimization. The comparative analysis of pressure drop variations for FD2-aligned configurations relative to both CS-45 and CS-90 is presented in Fig. 17 and Fig. 18. When compared to CS-45, Design 1 (D1) demonstrated average pressure drop reductions of 39 % and 56 % for inline and staggered arrangement respectively, while Design 2 (D2) showed configuration-dependent behavior: 4 % increase in the inline arrangement and 32 % decrease in the staggered arrangement. The dual-level configuration exhibited an 11 % increase in pressure drop.

In comparison with CS-90, D1 achieved reductions of 15 % and 39 % for the inline and staggered setups, respectively. D2 showed a 32 % increase in the inline arrangement but a 6 % decrease in the staggered one. The DL configuration showed a 54 % increase in pressure drop relative to CS-90.

Notably, the staggered configurations achieved the most substantial pressure drop reductions across all designs. These findings demonstrate that the proposed bio-inspired spacers can enhance mixing efficiency while simultaneously reducing pressure drop compared to conventional spacer designs, representing a significant advancement in spacer optimization.

Nonetheless, improving mixing remains the primary focus when implementing spacers, as the energy needed for pumping is comparatively low relative to the thermal energy requirements in MD processes, as highlighted by Woldemariam et al. [40].

Further analysis of the relationship between spacer void fraction (which are detailed in Table 1), and pressure drop, as shown in Fig. 19, that spacer orientation has a greater influence on pressure drop than void fraction. The results reveal that reducing the void fraction with a spacer orientation of $\theta = 0°$ causes the pressure drop to increase, whereas at $\theta = 90°$, the pressure drop decreases. This highlights that pressure drop at a given void fraction is strongly dependent on spacer orientation relative to the flow direction.

*3.3. Effect of the proposed bio-inspired spacers on membrane shear stress*

The effect of the most promising spacer configuration in this study, the FD2-aligned design, was evaluated by calculating the average shear stress along the membrane surface. Fig. 20 presents the results at an inlet velocity of 0.25 m/s. Compared to the empty channel case, the FD2-aligned spacer significantly increases the average shear stress on the bottom membrane surface and introduces greater spatial variation. This enhancement in shear stress contributes to reducing membrane fouling by limiting particle deposition and weakening the concentration polarization layer [39].

*3.4. Effect of the proposed bio-inspired spacers on temperature gradient*

The temperature contours presented in Fig. 21 illustrate the impact of the novel spacer configuration on the temperature distribution, in comparison to an empty channel, alongside the corresponding velocity streamlines. In the absence of spacers, the temperature profile exhibits a steep gradient near the membrane surface, accompanied by relatively uniform velocity streamlines, indicative of largely laminar flow conditions. This leads to pronounced temperature polarization, where the membrane surface temperature deviates significantly from the bulk fluid temperature. Conversely, the introduction of spacers disrupts the flow, generating localized turbulence and the formation of circulating vortices around the spacers. These vortices enhance convective mixing, which effectively reduces the thickness of the thermal boundary layer and improves heat transfer.

*3.5. Impact of the proposed bio-inspired spacers on mass and heat transfer*

Mass transfer characteristics within the feed channel, along with the effects of concentration polarization, are evaluated using the Sherwood number (*Sh*), a dimensionless parameter that represents the ratio of convective mass transfer to diffusive mass transfer (Eq. (3)). It serves as the mass transfer equivalent of the Nusselt number in heat transfer and is a crucial metric for assessing mass transfer efficiency [33].

Similarly, the thermal interaction within the hot/feed channel is analyzed using the Nusselt number (*Nu*), a dimensionless parameter, computed numerically using ANSYS, that represents the ratio of convective heat transfer to conductive heat transfer across the membrane surface. This parameter is vital for assessing heat exchange efficiency, as it quantifies the effectiveness of convective heat transfer.

To investigate the influence of geometry above the membrane surface on both mass and heat transfer performance, this study examines the effect of spacer orientation by analyzing two scenarios: (a) the regular scenario, where spacers are positioned on the membrane surface with their tips directed toward the feed channel wall (see Fig. 22a), and (b) the inverted scenario, where spacers are positioned on the feed channel wall with their tips directed toward the membrane surface (see Fig. 22b). These configurations are evaluated to determine their respective impacts on mass and heat transfer efficiency through the membrane surface.

The Sherwood number (*Sh*) was evaluated at various flow velocities in the region immediately adjacent to the membrane surface, with results depicted in Fig. 23. The analysis yielded several key insights. Spacers in the regular orientation consistently outperformed their inverted counterparts across all tested flow rates. Notably, the biomimetic spacer designs demonstrated enhanced performance relative to traditional spacer geometries, with particularly pronounced effects observed in the flow direction configuration (FD2) oriented at 90°. The highest improvement in Sherwood number compared to CS-45 and CS-90 was observed in the D1-FD2-Inline and D2-FD2-Inline configurations across all velocities examined under the standard conditions for spacers, as illustrated in Fig. 24, with increases of 15 % and 21 %, respectively, and at a flow velocity of 0.01 m/s.

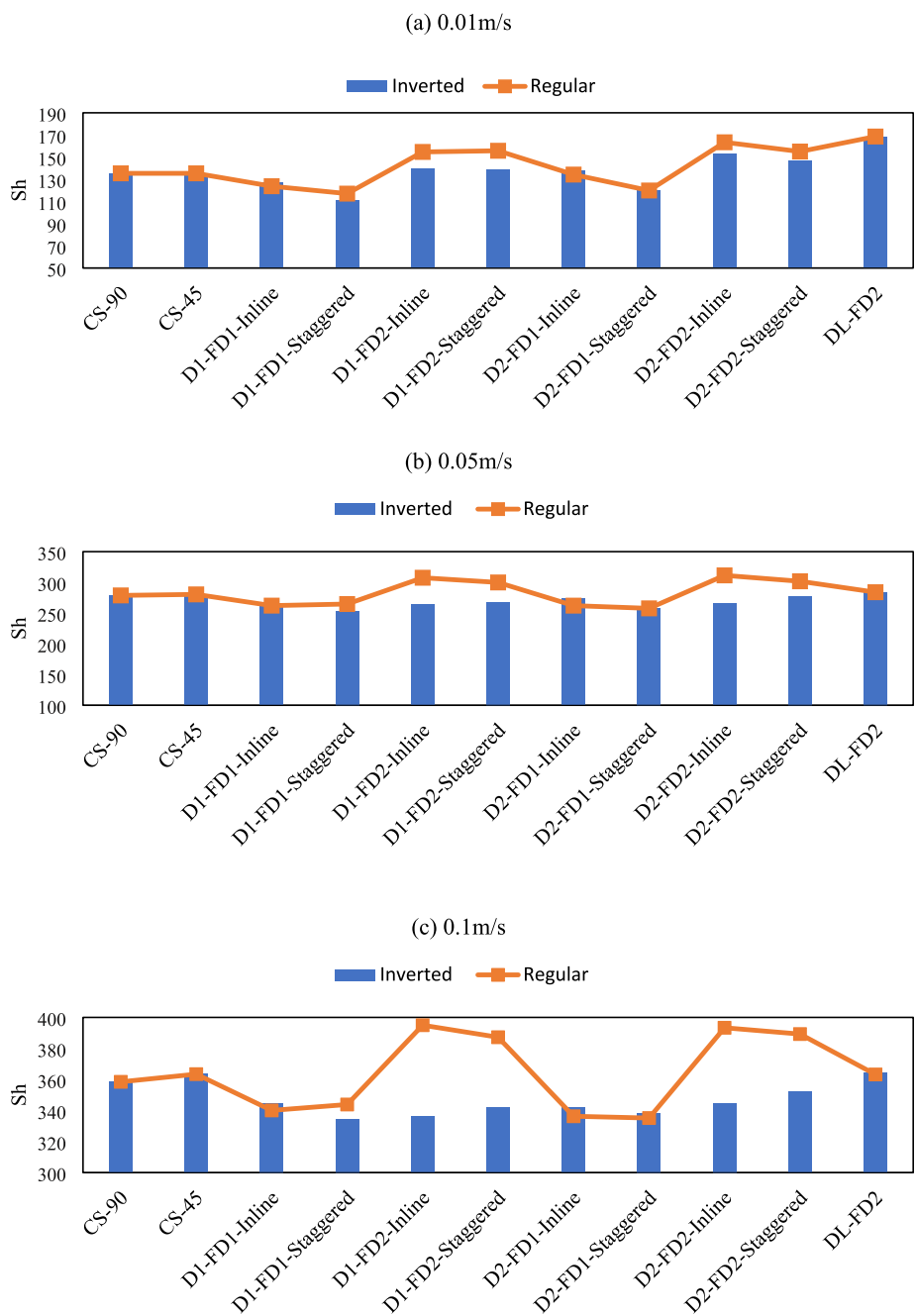

**Fig. 23.** Sherwood number (*Sh*) values for all proposed spacer configurations compared CS-90 and CS-45, analyzed in both the regular and inverted scenarios, across the velocity range studied.

From Fig. 25, it is observed that at the lowest flow velocity (Fig. 25a), the Nusselt number values in the regular scenario are slightly higher than those in the inverted scenario for most spacer configurations, with the highest increase of 5 % observed in the D2-FD2-Staggered configuration. Exceptions to this trend are the cylindrical and DL spacers, which exhibit symmetrical properties. At the highest flow velocity (Fig. 25b), the geometrical properties of the spacers have a more pronounced effect on the membrane, where the Nusselt number values for the regular configuration remain higher than those in the inverted configuration for most spacer designs. The largest increase, 17 %, is recorded in the D2-FD1-Staggered configuration, while the cylindrical and DL spacers again display symmetrical behavior. Overall, the data in Fig. 25 indicates that at low velocities, spacer configurations have a limited impact on Nusselt number values, whereas at higher velocities, the effect of spacer geometry becomes notable. Specifically, the flow direction configuration (FD2) at 90° achieves the highest Nusselt number values at the maximum velocity. Across both scenarios, the conventional spacer consistently achieves the highest Nusselt numbers overall. Thus, it can be concluded that the regular configuration outperforms the inverted

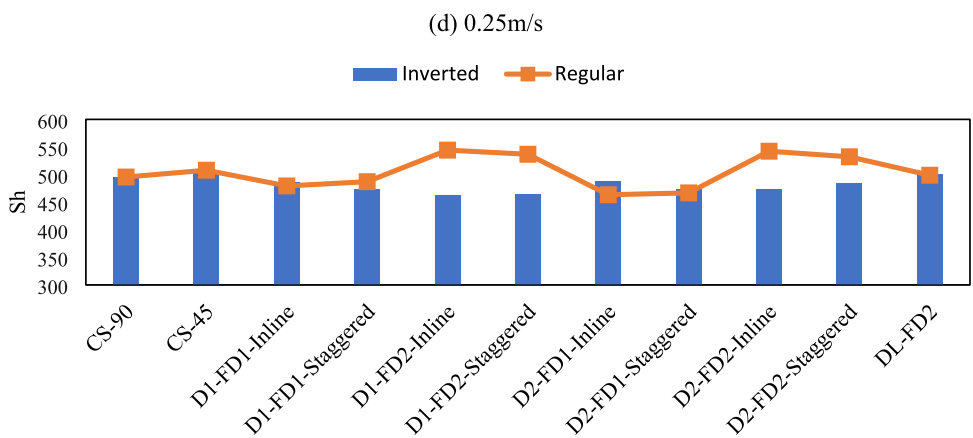

**Fig. 23.** (*continued*).

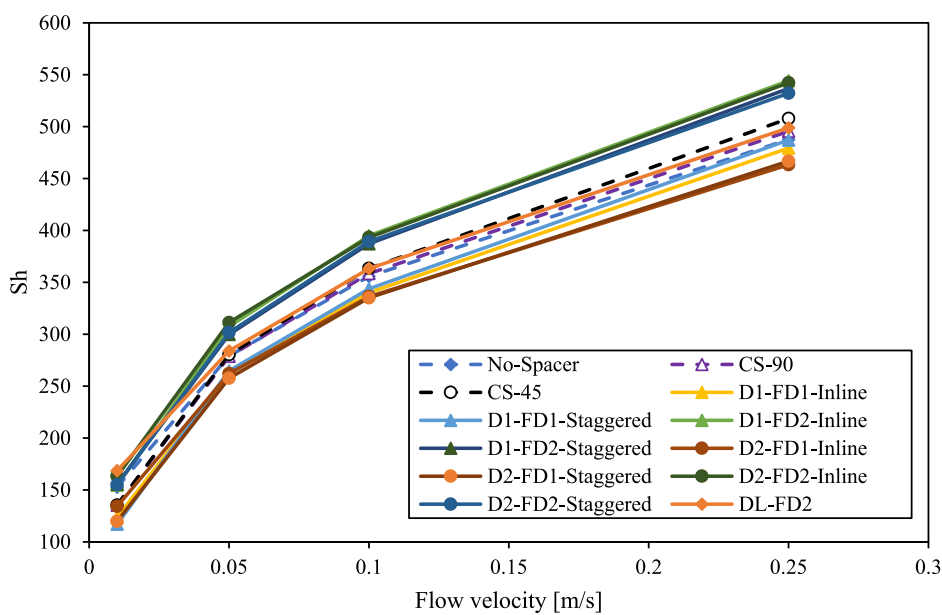

**Fig. 24.** Sherwood number (*Sh*) above the membrane wall for the regular scenario, comparing novel spacer configurations against the empty channel, CS-45 and CS-90 across 0.01–0.25 m/s inlet velocity.

configuration in terms of heat transfer efficiency.

## 4. Conclusions

This study presents a novel approach to spacer design for MD systems, drawing inspiration from the unique structure of alligator osteoderms. Through comprehensive 3D computational fluid dynamics (CFD) simulations, we successfully have demonstrated the potential of bio-inspired spacer configurations to enhance mixing efficiency in fluid systems. By examining various spacer designs and orientations, we identified configurations that significantly improve mixing performance while also addressing the associated challenges of increased pressure drop. The findings indicate that (DL-FD2) spacer arrangements can achieve up to a 5 times enhancement in mixing efficiency compared to traditional empty channel setups. This underscores the importance of optimizing spacer geometries not only to enhance mixing but also to minimize the energy required for pumping in applications such as membrane distillation.

The findings validate the considerable potential of bio-inspired designs in addressing critical challenges in membrane technology. By mimicking nature's time-tested solutions, we can develop more efficient and effective spacer geometries that significantly enhance the performance of MD systems. This approach not only improves mixing and reduces temperature polarization but also opens new avenues for innovation in membrane technology.

Overall, the proposed bioinspired spacers present a promising solution for improving fluid dynamics in various industrial processes, paving the way for further research into their implementation and long-term performance in real-world applications.

## CRediT authorship contribution statement

**Alaa Adel Ibrahim:** Writing – original draft, Visualization, Validation, Software, Methodology, Investigation, Conceptualization. **Stephan Leyer:** Writing – review & editing, Supervision, Investigation.

## Declaration of competing interest

The authors declare the following financial interests/personal relationships which may be considered as potential competing interests: Alaa Adel Ibrahim reports financial support was provided by FNR - Luxembourg National Research Fund. Reports a relationship with that includes:. Has patent pending to. If there are other authors, they declare that they have no known competing financial interests or personal relationships that could have appeared to influence the work reported in this paper.

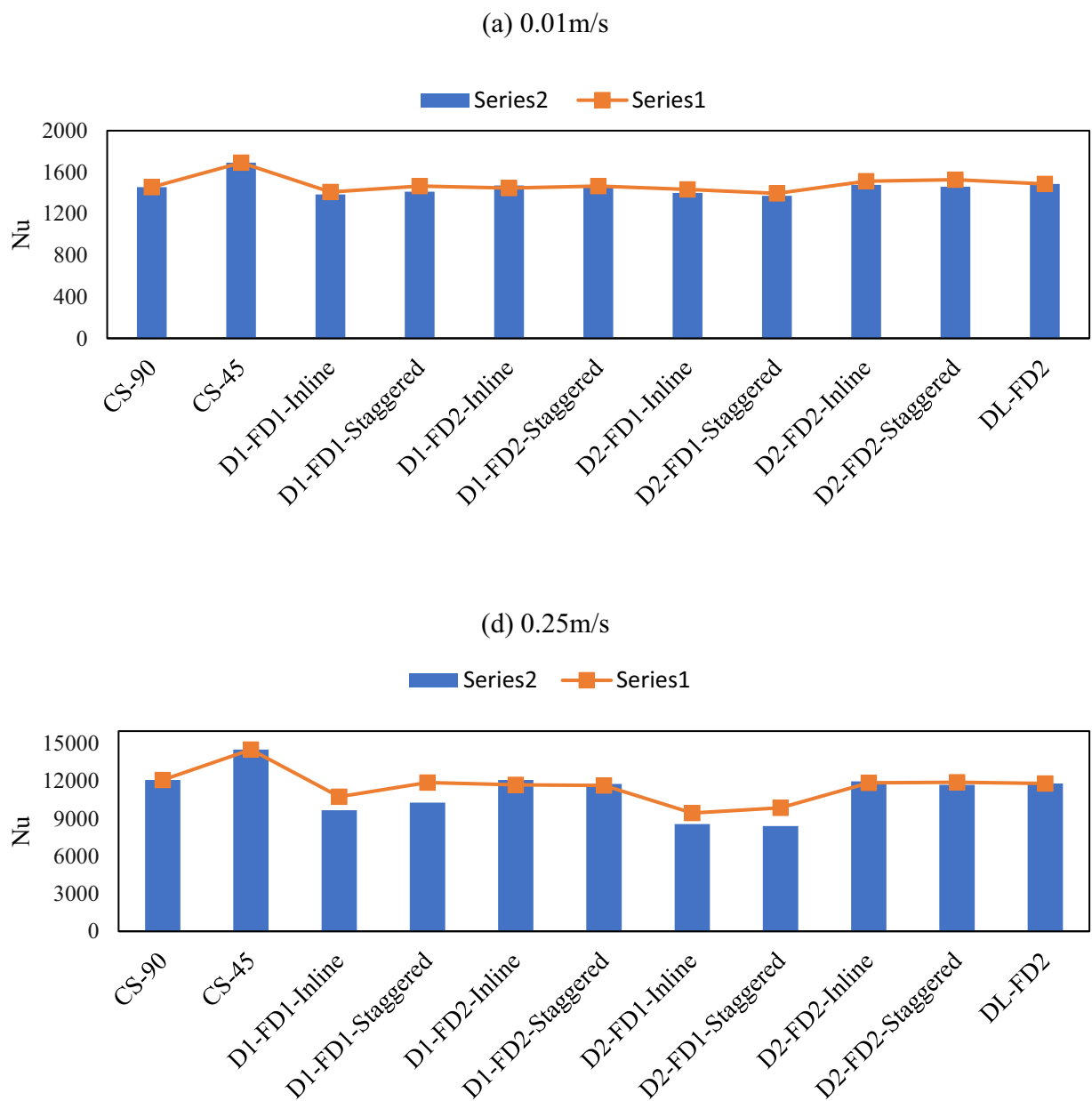

**Fig. 25.** Nusselt number values for all proposed spacer configurations compared to CS-90 and CS-45, analyzed in both the regular and inverted scenarios, across the velocity range studied.

**Acknowledgement**

The presented findings are part of the EXPCOAGMD project (AFR-17022904), supported by the Fonds National de la Recherche (FNR), Luxembourg.

**Data availability**

Data will be made available on request.